# Thermal Stability of Thermoelectric Materials via *In Situ* Resistivity Measurements

K. C. Lukas, W. S. Liu, Q. Jie, Z. F. Ren, C. P. Opeil
Department of Physics, Boston College, Chestnut Hill, MA, USA

ABSTRACT

An experimental setup for determining the electrical resistivity of several types of thermoelectric materials over the temperature range $20 < T < 550^\circ$ C is described in detail. One resistivity measurement during temperature cycling is also explained for $Cu_{0.01}Bi_2Te_{2.7}Se_{0.3}$ while a second measurement is made on $Yb_{0.35}Co_4Sb_{12}$ as a function of time at 400 $^\circ$C. Both measurements confirm that the materials are thermally stable for the temperature range and time period measured. Measurements made during temperature cycling show an irreversible decrease in the electrical resistivity of $Cu_{0.01}Bi_2Te_{2.7}Se_{0.3}$ when the measuring temperature exceeds the pressing temperature. Several other possible uses of such a system include but are not limited to studying the effects of annealing and/or oxidation as a function of both temperature and time.

INTRODUCTION

Thermoelectric materials have been widely studied over the last two decades with most of the research focused on increasing the dimensionless figure of merit ZT, $ZT=S^2T/\rho\kappa$, where S is the Seebeck coefficient, $\rho$ the electrical resistivity, $\kappa$ the thermal conductivity, and T the absolute temperature [1,2]. However ZT is not the only important parameter, especially when fabricating these materials for practical use, which is the ultimate goal. The materials should be relatively nontoxic, inexpensive, but most importantly their physical properties should remain stable over their temperature range of operation as well as during temperature cycling which most materials will experience in waste heat applications for cars [3], solar panel use [4], etc.

It is imperative when fabricating thermoelectric materials to ensure the materials are single phase and thermally stable. Much time and effort goes into the optimization of different synthesis parameters such as pressing temperature, annealing temperature, annealing time, etc. Several studies discussing the transport properties of thermoelectric materials also include thermal stability information based on temperature cycling, where the material is measured at different individual temperatures a number of times to see if the transport properties degrade after several runs. Another method to test the stability is to anneal the samples in a furnace at different temperatures for a varying amount of time and then measure the transport properties of the samples to study the effects of both the annealing time and temperature. Both methods are useful for basic information about the thermal stability, however both have shortcomings. Measurements taken at individual temperature intervals can miss important information such as phase transitions. And annealing samples in an oven at different temperatures for different times is useful, but the question of how many different temperature or time intervals should be established is difficult to answer. If there are not enough intervals, information may be missed or misinterpreted. If there are too many intervals then a great deal of time and/or sample preparation is required in order to obtain accurate information.

A solution to this problem can be made by measuring the transport properties *in situ* as the material is being annealed and/or cycled. The difficulty in decoupling the parameters in ZT [2] can now be seen as beneficial because by only measuring one transport property, accurate information about the material can be obtained as a function of both temperature and time; two important variables upon which thermal stability is dependent. We measure resistivity *in situ* as it is the most accurate measurement among the transport properties contributing to ZT. *In situ* resistivity measurements are commonly made on thin films to determine their temperature stability, but the authors have found no evidence or description of an experimental setup for *in situ* resistivity measurements on bulk thermoelectric materials to determine how properties change as a function of both annealing time and temperature. The following describes a setup to measure resistivity from room temperature to 550 $^o$C. It will be shown that the ability to control temperature as well as continuously read and collect data allows for much quicker and more accurate results for the determination of the thermal stability.

The paper is divided into two sections. The first describes in detail the experimental apparatus and configuration as well as comparisons with standard materials required to ensure an accurate determination of the resistivity as a function of temperature. The second section uses *in situ* measurements during temperature cycling to demonstrate how much more information can be ascertained quickly and easily on the thermal stability of a material, and finally other potential uses for the setup are stated but these data are not included.

EXPERIMENTAL DETAILS

The resistivity is setup and measured in a helium backfilled vacuum chamber (I). Samples, especially at high temperature must be measured in an oxygen free environment to prevent oxidation effects, unless that is the purpose of the measurement. Since the chamber must create an isolated environment, Viton o-rings are used to seal the chamber. The o-rings have a maximum operating temperature which when exceeded cause the o-rings to fail. In order to maintain an optimal operating temperature, a coolant plate (J) is used to dissipate excess heat. The coolant plate is kept at roughly 14 $^o$C via a closed loop water coolant system. This has proven to be enough cooling power to keep the o-rings under their maximum operating temperature, 100 $^o$C, while the inside of the chamber reaches temperatures in excess of 550 $^o$C. The chamber is evacuated with a mechanical pump down to pressures of 10 Pa. The chamber is then back filled with He gas, typically ambient pressure because it was determined that the cartridge heaters (B), which supply the heat, function better when in the presence of an exchange gas as opposed to in vacuum. Feed-throughs for electrical leads for current, voltage, heater power, and thermocouple inputs were mechanically fabricated. A sketch of the setup is shown in Figure 1. Nickel wire (3 mil) (H) is spark welded to each sample and then mechanically connected (D) to copper wire leads which in turn are mechanically connected to the vacuum chamber feed-throughs leading out of the chamber where the instrumentation for data acquisition are attached. Ni wire is used because it does not diffuse into the sample as readily as Cu, Au, or Ag which is of concern at high temperature. Temperature is read using 24 gage K-type thermocouple wire from Omega which is mechanically attached to the heating block with a screw; the 24 gage wire should be thick enough to negate any effects of "green rot" on the positive element which is a problem in oxygen depleted environments [5]. Mechanical connections are used at higher temperatures because solder or other electrically conducting epoxies are more difficult to use due to their lower operating temperatures. Heat is provided by a 120 V, 400 W cartridge heater (B) from Omega with a length of 3" and a 3/8" diameter. The cartridge heater is placed into a 1 x 1 x 3 inch stainless steel (SS) block (A) with a hole size slightly larger than the diameter of the cartridge heater.

In an ambient environment the cartridge heater resting in the SS block can typically reach 700 $^o$C. In vacuum this temperature is much more difficult to reach, and it was found that placing oxygen-free high thermal conductivity (OFHC) Cu shimstock inside the hole of the SS block creates greater surface contact area so that the SS block can remove the heat away from the cartridge heater, allowing the heater to reach higher temperatures without electrically shorting. Another necessity was to backfill the chamber with an exchange gas which can also transfer heat from the heater to the SS block. Combining these two effects allows the temperature to easily be raised to 550 $^o$C, and if necessary can reach up to 600 $^o$C. The sample (G) sits on top of a 0.1 mm thick layer of mica which is on top the of the SS block providing electrical insulation but the mica is also thin enough where it can be assumed that the sample temperature is the same as that of the heating block. To ensure the sample is thermally connected to the heating block, it is mechanically pressed down onto the block from above with a thin, 1/16" diameter, alumina rod (F) in a tungsten screw (C), which applies enough force to ensure good thermal contact but not enough to fracture the sample. The rod is thin and of low thermal conductivity which means heat flow out of the sample through the rod should be negligible.

The temperature is read and controlled by a PXR 4 (PID) temperature controller from Fuji Electric to which both the heater and K-type thermocouple are connected. The PXR 4 allows the rate at which the temperature is increased or decreased to be accurately controlled. Temperature is simultaneously read using a NI 9211 data acquisition system from National Instruments. The resistance is read using the 370 AC Resistance Bridge from LakeShore which uses an alternating current (AC) of 13.7 Hz. A LabVIEW program records the temperature (NI 9211), resistance (LS 370), time, and allows the user to set the frequency at which data is recorded. Unless otherwise noted data is recorded roughly once every second.

Resistivity measurements can be made either using a four point probe method on a bar shaped sample, or the Van der Pauw (VDP) technique [6]. The use of four probes negates any concerns about contact resistance [7-8]. The current used is AC with a frequency of 13.7 Hz which is sufficient to negate any voltage build up due to the Peltier and Seebeck effects [7]. The Peltier effect is due to the fact that when current flows from the current wire to the sample and out of the other current wire, heat is liberated at one junction and absorbed at the other due to the Peltier effect. This in turn will create a temperature gradient which will give rise to a voltage due to the Seebeck effect. When a direct current (DC) is used one must take this into consideration and switch the direction of the current to average out the excess voltage. However, the Peltier effect takes a finite amount of time to manifest itself which is usually on the order of 1 second [8]. So when an AC current is used, there is no concern of an excess voltage caused by thermoelectric effects.

For a bar shaped sample the resistivity is obtained from ρ=RA/L where R is the resistance, A is the cross sectional area, and L is the voltage lead separation. The placement of voltage leads always satisfies the ratio $2w < L_s - L$ where w is the thickness of the sample, $L_s$ is the length of the sample and L is the voltage lead separation which ensures uniformity of the electric field, or one dimensional current flow, at the voltage leads [7].

The Van der Pauw technique can be used to measure a sample of any arbitrary shape as long as the sample is flat and is singly connected, meaning it does not contain any holes [6,9]. The resistivity is given by the expression [6,9-10]

$$\rho = \frac{t\pi}{\ln(2)} \frac{(R_{12,34} + R_{23,41})}{2} F \qquad (1)$$

Where $R_{12,34}$ is defined as the current flowing between points 1 and 2 while the voltage is read between points 3 and 4 (inset Figure 1), $R_{23,41}$ has the current between points 2 and 3 with voltage read between 1 and 4, t is the thickness of the sample, and F is a correction factor that is a function of the ratio $R_r=R_{1234}/R_{2341}$ which can be solved graphically and is given by [10]

$$\frac{R_r - 1}{R_r + 1} = \frac{F}{\ln(2)} \text{arccosh}\left(\frac{\exp[\ln(2)/F]}{2}\right) \qquad (2)$$

Because thermoelectric materials have no widely accepted standard at high temperature (NIST only recently developed a low temperature standard [11]), it is imperative to accurately understand and account for any sources of error in the measurement so that data can be more accurately understood and communicated among research groups. The error bars for bar shaped samples from the propagation of independent errors are given by [12]

$$\frac{\sigma(\rho)}{\rho} = \sqrt{\left(\frac{\sigma(R)}{R}\right)^2 + \left(\frac{\sigma(L)}{L}\right)^2 + \left(\frac{\sigma(A)}{A}\right)^2} \qquad (3)$$

The error bars displayed for the Van der Pauw method are given by

$$\frac{\sigma_{VdP}(\rho)}{\rho} = \sqrt{\left(\frac{\sigma_{VdP}(R)}{R}\right)^2 + \left(\frac{\sigma(t)}{t}\right)^2} \qquad (4)$$

where

$$\sigma_{VdP}(R) = \sqrt{\sigma(R_{12,34})^2 + \sigma(R_{23,41})^2} \qquad (5)$$

It should be noted that Equation 1 is written under the assumption that the size of the contact points are infinitesimal and the contacts are made directly on the edge of the specimen. In reality the wire will always have some finite thickness and it is not possible to place the wire exactly on the edge of the sample, leading to additional error. This error is very difficult to quantify but should not be too large as long as care is taken in wire placement [13]. Therefore it is not taken into account in the expression for the error given in Equation 4, but should always be kept in mind.

Figures 2 and 3 show resistivity data while the temperature is increased in two different modes, discrete and continuous, for constantan and Ni respectively. Constantan is measured at discrete temperatures for a period of time on a bar shaped sample of dimensions 2 x 2 x 14 mm$^3$. The resistance values at each temperature are binned which gives the value of $\sigma(R)$ in Equation 3. It can be seen that the data measured by the constructed setup matches within 1% of the data taken by the ZEM-3 (ULVAC) on the standard constantan bar provided by ULVAC. The Ni data in Figure 3 is measured on a flat square shaped sample of dimensions 16 x 16 x 2 mm$^3$ using the Van der Pauw technique. The temperature was increased continuously from 20-550 $^{\circ}$C at a rate of 1 $^{\circ}$C/min to measure $R_{12,34}$. The sample was then cooled and wires reconfigured to measure $R_{23,41}$. The sample was again measured while the temperature was increased at 1 $^{\circ}$C/min. Resistance values were binned every degree to obtain the standard deviation. Though no error bars are expressed for the literature data, the agreement is within our experimental uncertainty up to temperatures of 375 $^{\circ}$C. The deviation at higher temperatures never exceeds 6%, but is in very good agreement. And the fact that it is not exact is not unexpected as it has been noted that the resistivity of Ni is very dependent on the sample purity [14]. The ferromagnetic transition temperature [15], which should not be as dependent on sample purity as the absolute resistivity value, is in very good agreement with the literature. The transition takes place at 355.5 $^{\circ}$C according to the literature [15], while the measurement here gives a transition temperature of 354 $^{\circ}$C which is well within the industrial error of 0.75% given for K-type thermocouples by Omega Engineering Inc.

Several other thermoelectric samples were run in order to validate the accuracy of the machine. These data are not included here as it would be redundant, however the results are summarized. For bar shaped samples where the same exact bar was measured in both the above setup as well as the ZEM-3, disagreement never exceeds 3% which is within the experimental uncertainty of the above system. The difference for Van der Pauw measurements never exceeds 9%, however if the ZEM is given an uncertainty of 3%, which is commonly used, there is again agreement within experimental error. There are several possible reasons for a greater disagreement in Van der Pauw measurements, but there are two that are most probable. The first is the fact that the resistivity is being compared between two different samples. One is a thin disk used for Van der Pauw measurements, while the other is a bar that is not cut from the same exact disk used for VP measurements; the ZEM can only measure bar shaped samples, and there can be slight variation among transport measurements of different samples of the same TE material. The second reason is due to the aforementioned effects of finite contact size and probe placement near the edge, so the difference noted above is not unexpected.

RESULTS

Now that the machine has been calibrated and benchmarked against standard samples as well as commercially available equipment, the setup can be used for its intended purpose of measuring resistivity during *in situ* annealing and/or temperature cycling. Figure 4 shows raw data for a $Cu_{0.01}Bi_2Te_{2.7}Se_{0.3}$ sample that has been temperature cycled. The sample was prepared via ball milling and dc hot pressing techniques described previously [16]; the sample in Figure 4 was hot pressed at 450 $^o$C. The temperature was incrementally cycled by running from 40 $^o$C up to 200 $^o$C at a rate of 5 $^o$C/min, then from 200 $^o$C back to 40 $^o$C at a rate of 5 $^o$C/min. The temperature was then ramped from 40 $^o$C to 225 $^o$C and then from 225 $^o$C back to 40 $^o$C. The system would remain at each maximum temperature for 10 minutes before cooling back down. This procedure was continued while increasing the maximum temperature each time by 25 $^o$C up to a temperature of 450 $^o$C. It can be seen that the sample exhibits metallic-like behavior. While several different runs were recorded, only the first run to a temperature of 200 $^o$C and the final run to 450 $^o$C are shown for clarity; the data from all intermediate temperatures lie in between the warming curves for both 200 and 450 $^o$C. The values for resistivity change by less than 5% while the temperature remains below the pressing temperature. However once the pressing temperature is reached, and slightly exceeded as the temperature always overshoots the set value by a few degrees due to the fast ramp rate, the resistivity value is lowered by about 13% from its initial value. The dc hot press method essentially anneals the sample at the pressing temperature, and if the pressing temperature is exceeded during measurement or operation, there are irreversible changes to the transport properties of the material due to further annealing. While both the Seebeck coefficient and thermal conductivity values are just as important as resistivity values, it is evident how quickly information can be ascertained from the continuous measurement capability and how useful it can be in acquiring further information.

Another possibility that is easy to realize besides temperature cycling is *in situ* annealing measurements as a function of time, which has been measured for $Yb_{0.35}Co_4Sb_{12}$ and is shown in Figure 5. The sample can be brought to a specific temperature and remain at that temperature to study the effects of annealing or operating temperature over a period of time. The typical range of operation for skudderudites is between 400 – 500 $^o$C for waste heat applications [8]. The sample shown in Figure 5 was heated from 20 to 400 $^o$C at a rate of 5 $^o$C/min. The temperature was then held at 400 $^o$C for 36 hours. Then the temperature was lowered from 400 to 20 $^o$C again at a rate of 5 $^o$C/min. As mentioned previously the time interval at which data is recorded can be changed. During warming and cooling the data was recorded roughly every second. To minimize the number of data points during the 36 hour period, data was recorded every 10 minutes. The frequency of data acquisition can be set to as long or as short as desired. Figure 5a shows raw data for how the temperature (bottom) and resistivity (top) vary with time, while Figure 5b shows ρ plotted against temperature. From Figure 5a we see that the ramp rate is constant on the way up as expected and the temperature is stable at 400 $^{o}$C for the entire 36 hours. The resistivity at 400 $^{o}$C changes by less than 1 % over the 36 hour period. The cooling rate is constant until a temperature of 80 $^{o}$C is reached where the system does not have the ability to cool at the 5 $^{o}$C/min through convective and conductive cooling. From Figure 5b it is seen that the resistivity shows no hysteresis even with the different cooling rates below 80 $^{o}$C and the difference in ρ after heating is about 1 %. It is the intention of this work to demonstrate the capabilities of the apparatus, and not to provide an in depth analysis on the thermal stability of $Cu_{0.01}Bi_2Te_{2.7}Se_{0.3}$ and $Yb_{0.35}Co_4Sb_{12}$, though that work is presently underway, and so there will be no further discussion of the results. Though it should be noted that bismuth telluride and skudderudite compounds are not thermally stable at higher temperatures or longer operation times than those reported in this work.

Only brief mention will be made of yet another capability for the setup which would be to study oxidation effects on thermoelectric materials. It was stated previously that the chamber is either evacuated or backfilled with an inert gas. It is possible to leave air in the chamber to study the effects of oxygen with both time and temperature. No data is presented here but it is straightforward to see how these measurements can readily be made. From the ability to make *in situ* resistivity measurements along with the strong interrelation between the components of ZT, the benefits of this setup become immediately apparent. The ability to measure ρ continuously as the temperature is being cycled and/or held constant leads to much more information than would be obtained with simple incremental measurements as is the case in phase transitions, over several temperature cycles, or while annealing as was demonstrated above in the measurements of $Cu_{0.01}Bi_2Te_{2.7}Se_{0.3}$ and $Yb_{0.35}Co_4Sb_{12}$.

CONCLUSION

An experimental setup for accurately determining the electrical resistivity of thermoelectric materials in a temperature range of 20-550 $^{o}$C has been described in detail. Two *in situ* resistivity measurements have also been described. Measurements were made on $Cu_{0.01}Bi_2Te_{2.7}Se_{0.3}$ hot pressed at 450 $^{o}$C while the temperature was cycled, and from the data it was determined that the material is stable upon cycling while the temperature does not exceed the pressing temperature. Measurements were also made on $Yb_{0.35}Co_4Sb_{12}$ held at its operating temperature for 36 hours. The material is again stable over this time period. Other possible measurements including the study of annealing as well as oxidation effects can be easily realized with the apparatus.

ACKNOWLEDGEMENTS

The authors gratefully acknowledge funding for this work through the "Solid State Solar-Thermal Energy Conversion Center (S3TEC)", an Energy Frontier Research Center

List of Captions

Figure 1: Diagram of experimental setup for in situ resistivity measurements. Alumina rods, ¼", (F) are connected to SS pieces (E) machined with hole sizes a fraction larger than the alumina rods. The inset shows the wiring configuration for a sample being measured using the Van der Pauw technique.

Figure 2: The percent error is plotted with data obtained from measurements made on the same standard bar shaped sample using a standard four point probe (SFPP) technique in the setup shown in Figure 1 as well as the commercially available ZEM-3. The inset shows resistivity of constantan plotted versus temperature along with uncertainty calculated from Equation 3 demonstrating agreement between both systems with the standard.

Figure 3: Resistivity of nickel is plotted against temperature. Measurements were made using the Van der Pauw technique on a sample of 99.9993% purity from AJA International, Inc. Values obtained from ref. [13] for Ni of 99.98% purity are shown for comparison.

Figure 4: Resistivity is plotted against temperature for $Cu_{0.01}Bi_2Te_{2.7}Se_{0.3}$ hot pressed at $450^o$ C during temperature cycling. Negligible change is seen in the material while the temperature remains below the pressing temperature.

Figure 5: Resistivity and temperature are plotted against time (a). The resistivity is plotted with temperature (b) where there is negligible difference in the warming and cooling curves, even at different cooling rates.

Figure 1

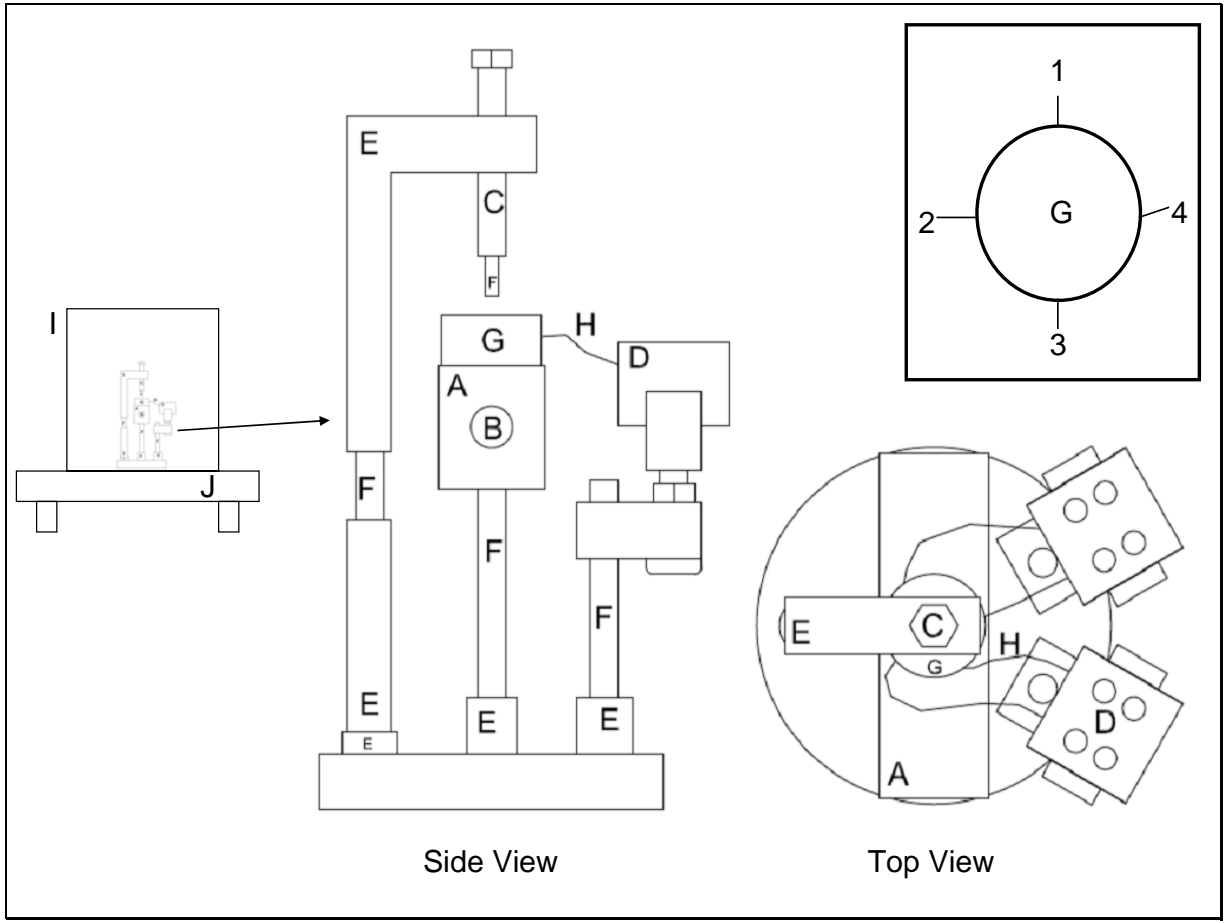

Figure 2

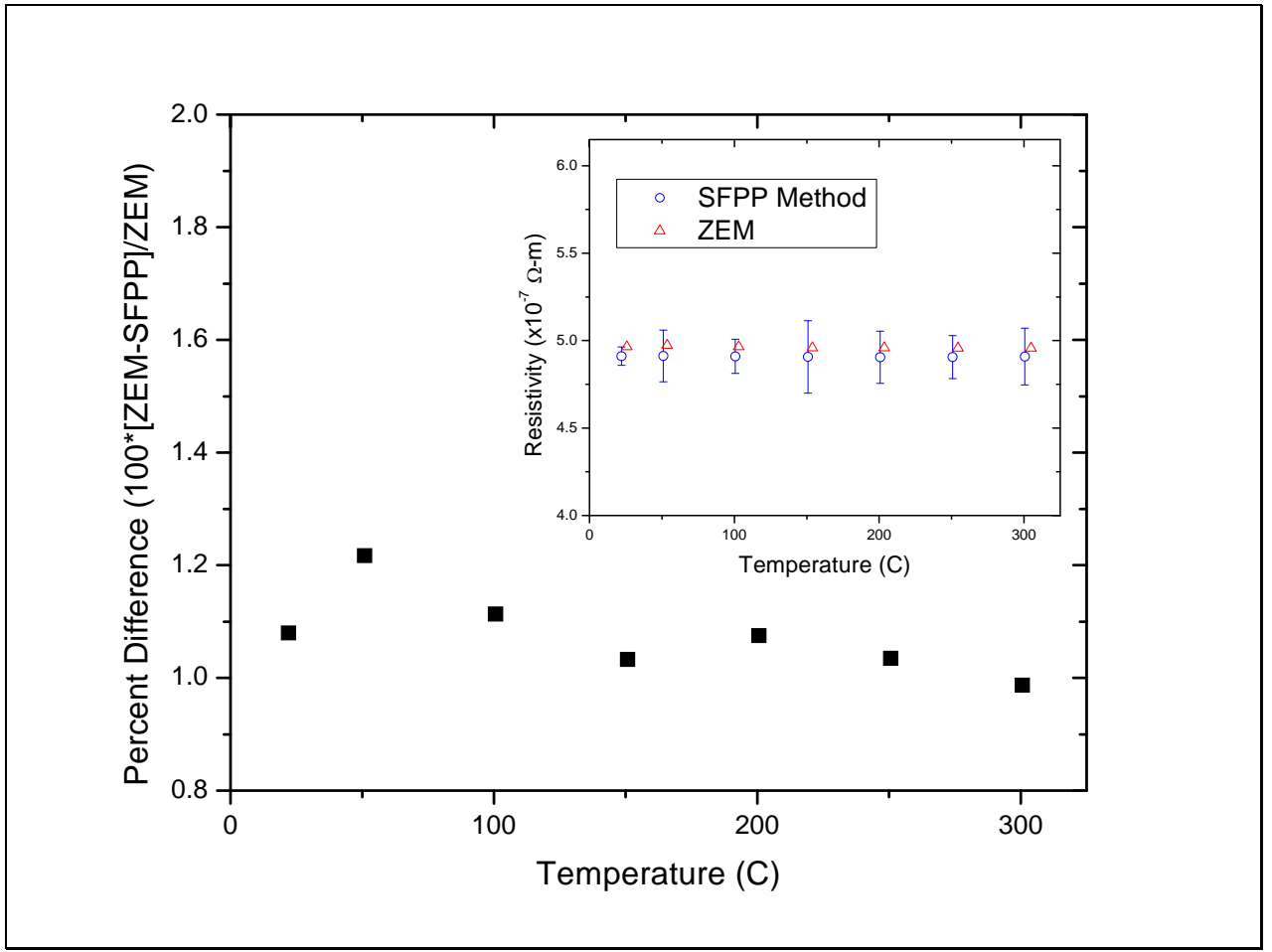

Figure 3

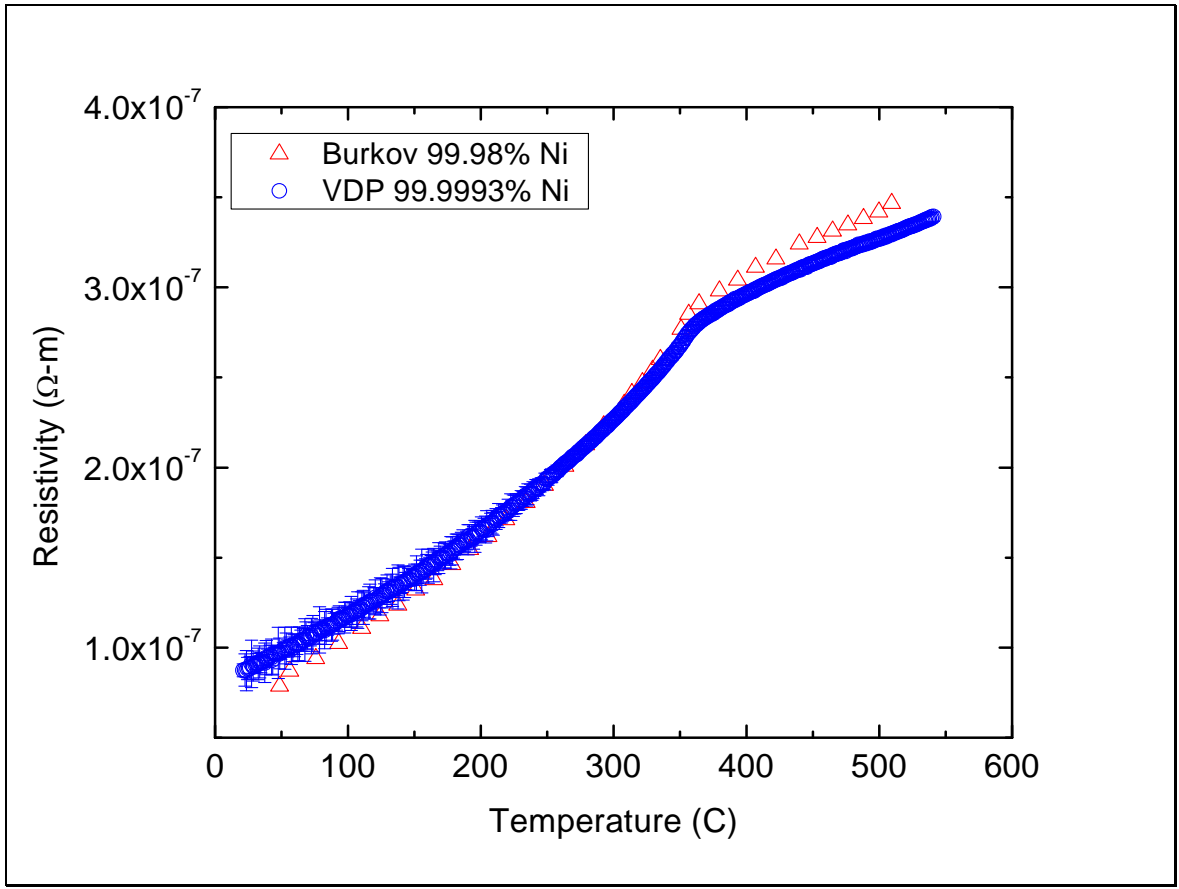

Figure 4

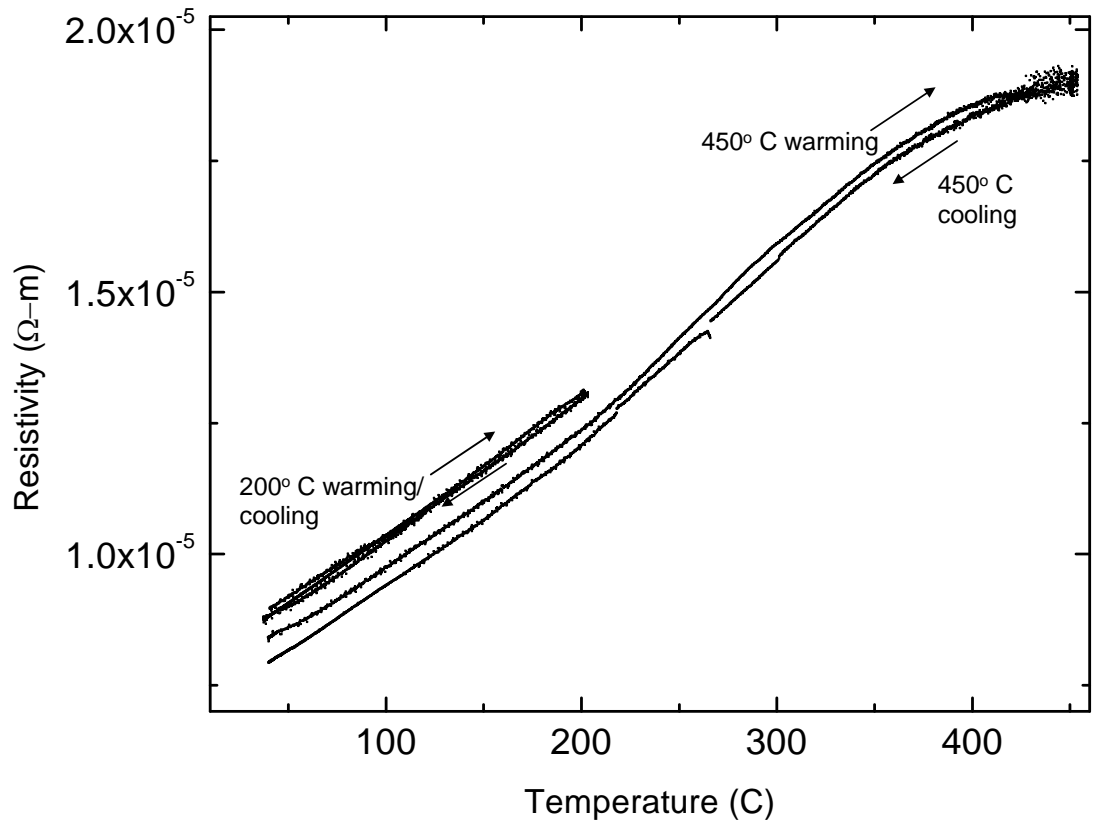

Figure 5

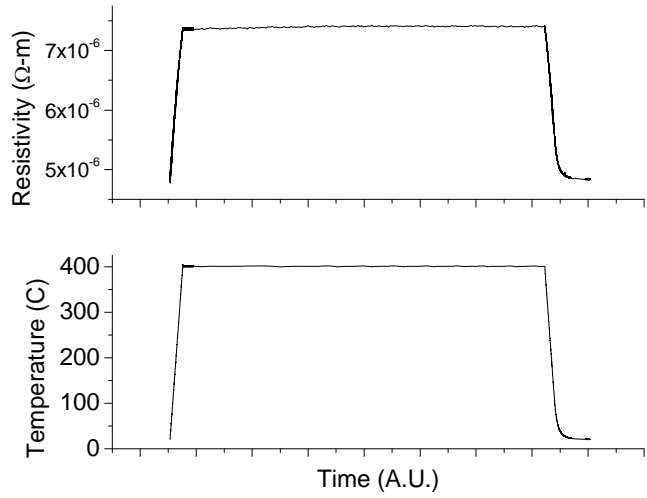 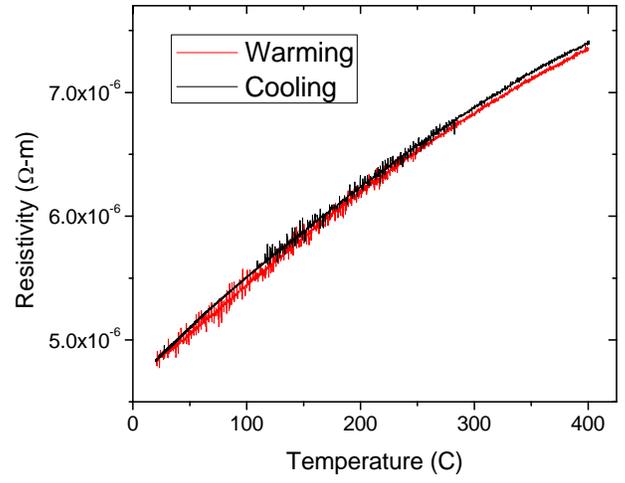

a

b